\documentclass[prb,preprint,showpacs,superscriptaddress,unsortedaddress]{revtex4}
\usepackage{amsfonts}
\usepackage{amsmath}
\usepackage{amssymb}
\usepackage{graphicx}
\usepackage{dcolumn}
\usepackage{bm}
\usepackage{color}
\usepackage{makeidx}
\usepackage{ulem}
\usepackage{hyperref}
\begin{document}

\title{Occurrence of a high-temperature superconducting phase in
 cuprate/titanate (CaCuO$_2$)$_n$/(SrTiO$_3$)$_m$ superlattices}

\author{D. Di Castro}
\affiliation{CNR-SPIN and Dipartimento di Ingegneria Civile e
Ingegneria Informatica, Universit\`a di Roma Tor Vergata, Via del
Politecnico 1, I-00133 Roma, Italy}
\author{M. Salvato}
\affiliation{Dipartimento di Fisica and MINAS Lab., Universit\`a
di Roma Tor Vergata, I-00133 Roma, Italy}
\author{A. Tebano}
\affiliation{CNR-SPIN and Dipartimento di Ingegneria Civile e
Ingegneria Informatica, Universit\`a di Roma Tor Vergata, Via del
Politecnico 1, I-00133 Roma, Italy}
\author{D. Innocenti}
\affiliation{CNR-SPIN and Dipartimento di Ingegneria Civile e
Ingegneria Informatica, Universit\`a di Roma Tor Vergata, Via del
Politecnico 1, I-00133 Roma, Italy}
\author{C. Aruta}
\affiliation{CNR-SPIN, Dipartimento di Scienze Fisiche, Via
Cintia, Monte S.Angelo, 80126 Napoli, Italy }
\author{W. Prellier}
\affiliation{Laboratoire CRISMAT, UMR 6508, CNRS-ENSICAEN 6Bd
Marechal Juin, 14050 Caen, France}
\author{O. I. Lebedev}
\affiliation{Laboratoire CRISMAT, UMR 6508, CNRS-ENSICAEN 6Bd
Marechal Juin, 14050 Caen, France}
\author{I. Ottaviani}
\affiliation{Dipartimento di Fisica and MINAS Lab., Universit\`a
di Roma Tor Vergata, I-00133 Roma, Italy}
\author{N. B. Brookes}
\affiliation{European Synchrotron Radiation Facility, 6 rue Jules
Horowitz, BP 220, 38043 Grenoble, Cedex 9, France}
\author{M. Minola}
\affiliation{CNISM and Dipartimento di Fisica, Politecnico di
Milano, I-20133, Italy}
\author{M. Moretti Sala}
\affiliation{CNISM and Dipartimento di Fisica, Politecnico di
Milano, I-20133, Italy}
\affiliation{European Synchrotron Radiation Facility, 6 rue Jules
Horowitz, BP 220, 38043 Grenoble, Cedex 9, France}
\author{C. Mazzoli}
\affiliation{CNISM and Dipartimento di Fisica, Politecnico di
Milano, I-20133, Italy}
\author{P.G. Medaglia}
\affiliation{CNR-SPIN and Dipartimento di Ingegneria Industriale,
Universit\`a di Roma Tor Vergata, Via del Politecnico 1, I-00133
Roma, Italy}
\author{G. Ghiringhelli}
\affiliation{CNISM and Dipartimento di Fisica, Politecnico di
Milano, I-20133, Italy}
\author{L. Braicovich}
\affiliation{CNISM and Dipartimento di Fisica, Politecnico di
Milano, I-20133, Italy}
\author{M. Cirillo}
\affiliation{Dipartimento di Fisica and MINAS Lab., Universit\`a
di Roma Tor Vergata, I-00133 Roma, Italy}
\author{G.Balestrino}
\affiliation{CNR-SPIN and Dipartimento di Ingegneria Civile e
Ingegneria Informatica, Universit\`a di Roma Tor Vergata, Via del
Politecnico 1, I-00133 Roma, Italy}

\begin{abstract}
 We  report the occurrence of superconductivity, with maximum
\textit{T$_{c}$} = 40 K, in superlattices (SLs) based on two
insulating oxides, namely CaCuO$_{2}$  and SrTiO$_{3}$. In these
(CaCuO$_{2}$)$_n$/(SrTiO$_{3}$)$_m$ SLs, the CuO$_{2}$ planes
belong only to CaCuO$_{2}$ block, which is an antiferromagnetic
insulator. Superconductivity, confined within few unit cells at
the CaCuO$_{2}$/SrTiO$_{3}$ interface, shows up only when the SLs
are grown in a highly oxidizing atmosphere, because of extra
oxygen ions entering at the interfaces. The hole doping  is
obtained by charge transfer from the interface layers, which thus
act as charge reservoir.
\end{abstract}

\date{\today }
\maketitle

\section{Introduction} The parent compounds of high transition temperature
(\textit{T$_{c}$}) cuprate superconductors (HTS) are
antiferromagnetic insulators and develop superconductivity when
charge carriers are introduced in the CuO$_{2}$ planes; this
injection is usually achieved by charge transfer from a properly
doped structural subunit (charge reservoir). A different approach
to obtain high \textit{T$_{c}$} superconductivity is based on the
engineering of heterostructures (HSs) or superlattices (SLs) by
advanced layer by layer deposition techniques. Up to now this
approach has been applied to artificial structures consisting of
an insulating and a metallic cuprate, such as
CaCuO$_{2}$/BaCuO$_{2}$ \cite{Bales1,Tebano} and
La$_{2}$CuO$_{4}$/La$_{2-x}$Sr$_{x}$CuO$_{4}$ (LCO/LSCO).
\cite{Gozar} In these systems the metallic cuprate acts as charge
reservoir, injecting carriers in the CuO$_2$ planes of the
insulating cuprate, thus artificially reproducing the mechanism
which naturally occurs in cuprate HTS.

Bearing in mind this mechanism one could speculate about different
possible choices of the charge reservoir block oxide, not
necessarily a metallic cuprate. On the other hand, it has been
shown that, by an adequate choice of the two constituents, it is
possible to realize a metallic interface, which can be itself
superconducting, as in the case of the oxide
LaAlO$_{3}$/SrTiO$_{3}$ HS, \cite{Reyren} semiconducting
monochalcogenide HSs, \cite{Fogel} bicrystals of semimetals
\cite{Muntyanu} and multipgraphene. \cite{Barzola}

This extraordinary interface phenomenon suggests that it could be
possible to identify two systems, one copper oxide and one
copper-free oxide, in order to obtain a  doped interface, which
could act as charge reservoir for the cuprate block.
  Following this idea we
have chosen two insulating oxides,  the SrTiO$_{3}$ (STO) and the
CaCuO$_{2}$ (CCO), as building blocks for the engineering of
(CCO)\textit{$_{n}$}/(STO)\textit{$_{m}$} SLs, where \textit{n}
and \textit{m} are the number of unit cells (u.c.) of CCO and STO,
respectively. SrTiO$_{3}$ is at the basis of the emerging field of
oxide electronics because of the occurrence of exotic,
two-dimensional phases of electron matter at the interface with
other oxides. \cite{Reyren,Cen,Dagotto,Nakagawa} CaCuO$_{2}$ is an
antiferromagnetic insulator \cite{Sigrist} and it is considered
the simplest parent compound of HTS \cite{Azuma}: in this compound
the CuO$_{2}$ planes, where superconductivity can occur, are
separated by bare Ca atoms in a pure infinite layer (IL)
structure. \cite{Sigrist}
In some IL alloys, obtained by chemical synthesis under high
pressure (several GPa) conditions, superconductivity  was reported
\cite{Azuma,Zhang} and explained by the random insertion of planar
defects within the IL structure. \cite{Zhang,Tao}

In this work, we provide evidences that a high critical
temperature superconducting phase, with a maximum \textit{T$_{c}$}
= 40 K, can be  obtained in a controlled way by using the IL CCO
and the model system STO as building blocks for the synthesis of
(CCO)\textit{$_{n}$}/(STO)\textit{$_{m}$} SL films. In these
systems the superconductivity develops at the interfaces between
the two constituent oxides because of the peculiar structural
characteristics of the CCO/STO interfaces. These, indeed, are
hybrid between the perovskite and the infinite layer structure,
and thus can leave space for extra oxygen ions to come in and dope
the system. The charge reservoir is  likely provided by the
interface layers, which inject holes in the inner CuO$_2$ planes
of the CCO block, making the SLs superconducting, although
constituted by two insulating materials.  The superconducting
properties of these SLs can be controlled and systematically
investigated by varying the growth conditions and the relative
thickness of the constituent blocks.

\section{EXPERIMENTAL DETAILS}

\subsection{Sample growth}

We used the pulsed laser deposition technique (KrF excimer laser =
248 nm) to synthesize several superlattice
(CCO)\textit{$_{n}$}/(STO)\textit{$_{m}$} films, made by 10 to 20
repetitions of the supercell constituted by \textit{n} u.c. of CCO
and \textit{m} u.c. of STO. The films were deposited  on 5x5
mm$^2$ NdGaO$_{3}$ (110) (NGO) oriented mono-crystalline
substrates obtained from Crystal, GmbH. NGO is the most suitable
substrate to grow both CCO (\onlinecite{Bales3}) and
 STO, having a pseudocubic in plane lattice parameter (a = 3.87 \AA) just in the middle between CCO (a =
3.84 \AA) and STO (a = 3.91 \AA). Two targets, with CaCuO$_{2}$
and SrTiO$_{3}$  nominal composition, mounted on a multitarget
system, were used. The STO target is a commercial crystal obtained
from Crystal, GmbH. The CCO target was prepared by standard solid
state reaction, according to the following procedure:
stoichiometric mixtures of high-purity CaCO$_3$ and CuO powders
were calcined at 860$^{\circ}$C in air for 24 h, pressed to form a
disk, and finally heated at 900$^{\circ}$C for 12 h. The substrate
was placed at a distance 2.5 cm from the targets on a heated
holder and its temperature during the deposition of the
 SLs was $T \simeq$ 600$^{\circ}$C. For the growth
of the superconducting SLs the deposition chamber was first
evacuated down to P $\sim$ 10$^{-5}$ mbar and then a mixture of
oxygen and 12\% ozone atmosphere at a pressure of about 1 mbar was
used, followed by a rapid quenching in high oxygen pressure (about
1 bar). Non-superconducting SLs were grown at the same temperature
and pressure but with no ozone and no high pressure quenching, as
explained in the subsequent sections of the paper. NdO surface
terminated NGO substrates,   obtained by annealing in air at
1000$^{\circ}$C for 2h, \cite{Ohnishia} were also used. We found
that there is no
 appreciable difference among samples grown on NGO with controlled and uncontrolled termination
 from the point of view of structural quality (as evinced from x-ray diffraction measurements) and superconducting properties.

The deposition rate of the two constituent blocks was first
estimated by measuring ex-situ the thickness of single phase films
of CCO and STO, grown on NGO substrate at the same growth
conditions used for the SLs. The thicknesses  were obtained by
finite size oscillations analysis of the x-ray diffraction (XRD)
spectra of the films. Afterwards, an accurate estimate of the
thickness of CCO and STO within the superlattice structure was
made by the analysis of the XRD spectra of the SLs. Indeed, in the
XRD spectrum of a system with SL structure, as the
(CCO)\textit{$_{n}$}/(STO)\textit{$_{m}$}, satellite peaks
SL$_{+1}$ and SL$_{-1}$ show up around the average structure peak
SL$_{0}$ (see Fig.\ref{structure}a). From the positions
2$\theta_{+1}$ and 2$\theta_{-1}$ of the satellite peaks (see
Fig.\ref{structure}a), it is possible to evaluate accurately the
thickness of the supercell $\Lambda =$ \textit{n} $\cdot$
c$_{CCO}$ + \textit{m} $\cdot$ c$_{STO}$ (c$_{CCO}$ and c$_{STO}$
are the c-axis lattice parameters of CCO and STO, respectively),
by using the following formula\cite{Bales1,Schuller}: $\Lambda =
\lambda/[sin(\theta_{+1}) - sin(\theta_{-1})]$, were $\lambda$ is
the wavelength of the incident radiation. From the position
2$\theta_0$ of the average structure peak SL$_0$, the value of the
mean c-axis lattice parameter c = ( \textit{n} $\cdot$ c$_{CCO}$ +
\textit{m} $\cdot$ c$_{STO}$)/(\textit{n+m}) is easily obtained
by: c = $\lambda$/ [2sin($\theta_0$)]. To estimate the individual
values of \textit{n} and \textit{m}, we synthesized several SLs
(CCO)\textit{$_{n}$}/(STO)\textit{$_{m}$} keeping fixed the number
of laser shots on the STO target and varying the number $N$ of
laser shots on the CCO target.
\begin{figure}[tbp]
\vspace{0.5cm}
\par
\begin{center}
\includegraphics[width=7cm]{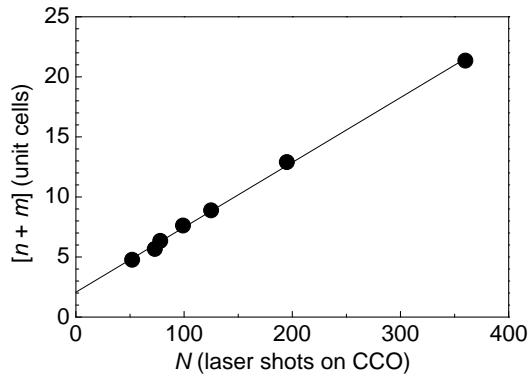}
\end{center}
\caption{Thickness of the supercell ($n + m$), in unit cells, as a
function of the number $N$ of laser shots on the CCO target. The
extrapolation to $N = 0$ gives the number of unit cells of STO
block.} \label{n+m}
\end{figure}
In Fig. \ref{n+m} we report   $\Lambda/$c = \textit{n+m},
estimated by the XRD spectra as described above, as a function of
the laser shots $N$ on the CCO target for various
(CCO)\textit{$_{n}$}/(STO)\textit{$_{m}$} SLs. The value of
\textit{m} is kept fixed at about 2, as evaluated by the rough
estimate of the deposition rate from the XRD spectra of the single
phase films. The experimental points lie on a straight line, as
expected if all the growth conditions (pressure, temperature,
laser fluence, target to substrate distance) are not changed from
one deposition to another. The extrapolation to $N = 0$ gives the
thickness \textit{m} (in unit cells) of the STO layer and,
consequently, the thicknesses of the CCO layer and the deposition
rate of both STO and CCO. In our case, from a linear fit we
obtained (approximating to half cell) \textit{m} = 2, and thus
\textit{n} = 3, 3.5, 4, 5.5, 7, 10.5, 19.5.

\subsection{Experimental techniques}
 The structural characterization was performed by x-ray diffraction measurements in a  $\theta-2\theta$
   Bragg-Brentano geometry and by high-resolution transmission electron microscopy (HRTEM). The TEM images were taken
   using a Tecnai G2 30 UT microscope with a 1.7 \AA point resolution, operated at 300 kV.
   Cross-section specimens for HRTEM were prepared by mechanically grinding down to the thickness
   of approximately 20 $\mu$m, followed by Ar+ ion beam milling using a Ion Slicer EM-09100 IS apparatus.
   The HRTEM image simulation has been done using Mac Tempas and Cristal Kit software.
   A Bragg-mask filter was used in order to enhance the noise/signal ratio.

\begin{figure}[tbp]
\vspace{0.5cm}
\par
\begin{center}
\includegraphics[width=5cm]{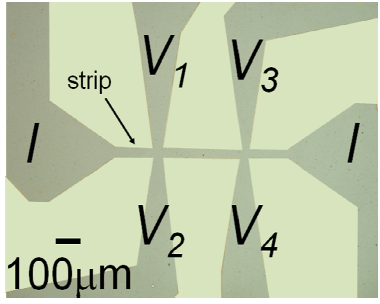}
\end{center}
\caption{(color online) Representative
 picture of a sample patterned with standard  lithographic process.
 The darker  area corresponds to the sample surface. The position of the current and voltage  terminals
 are indicated with \textit{I}  and \textit{V}, respectively, whereas the black arrow indicate the  strip.} \label{patterned}
\end{figure}

The samples for resistivity, \textit{I-V} characteristics, and Hall effect measurements
were patterned using a standard  lithographic process in order to
obtain a definite geometry. In Fig.\ref{patterned} a
representative
 picture of the patterned samples  is shown. In this image, the darker  area corresponds to the sample surface with the
 horizontal thin strip and the position of the current \textit{I} and voltage \textit{V} terminals clearly visible.

In the case of the \textit{I-V} measurements  the strip width was
w = 90 $\mu$m  and the distance between the in-line voltage
contacts was L = 400 $\mu$m. No substantial changes in the
\textit{I-V}  characteristics and resistivity values were observed
for samples patterned with strip widths in the range
 50-100 $\mu$m. The resistivity and \textit{I-V} measurements were performed using the standard
 four-probe dc
 technique, where two of the in line V terminals (V$_1$ and V$_3$ or alternatively V$_2$ and V$_4$) were used as
 voltage probes. For all the measurements, the current electrodes were attached at the left and right \textit{I} labelled
 terminals and the current has been considered uniformly
 distributed along the sample thickness.

The Hall voltage $V_H$ was measured between two of the opposite
voltage terminals ($V_{H^{1-2}} = V_1 - V_2$ or
 alternatively $V_{H^{3-4}} = V_3 - V_4$) with the external magnetic field $B$ perpendicular to the sample surface. In this
 case the strip width was 50 $\mu$m while the distance between two pairs of the in-line terminals was the
 same as for \textit{I-V} measurements. During Hall voltage measurements, in order to reduce possible spurious
 effects due to sample and strip uniformity, the directions of the current and the magnetic field were
 inverted and $V_H$ was measured between both the opposite voltage pairs for each value of $B$ and $I$. The
 difference in the measured values of $V_H$ in the different configurations resulted to be less than the
 experimental uncertainty.

X-ray absorption spectroscopy (XAS) experiment was performed at
the beamline ID08 of the European Synchrotron Radiation
 Facility using the high scanning speed Dragon-type monochromator. The x-ray source was an Apple II
 undulator delivering almost 100\% polarized radiation in both horizontal and vertical directions. The
 total electron yield detection had a probing depth of 3-6 nm, so that several superlattice cells under the
 surface could be investigated. The incident beam was forming a 60 degrees angle with the sample
 surface normal, which is parallel to the crystal c-axis. In this geometry, with vertical polarization the
 electric field vector $\textbf{E}$ of the incident radiation was parallel to the $ab$-plane; with horizontal polarization
 $\textbf{E}$ was mostly parallel to the c axis, so that 75\% of the intensity comes from the final states lying perpendicular
 to the $ab$-plane.

\section{Results and discussion}
\begin{figure*}[tbp]
\vspace{0.5cm}
\par
\begin{center}
\includegraphics[width=10cm]{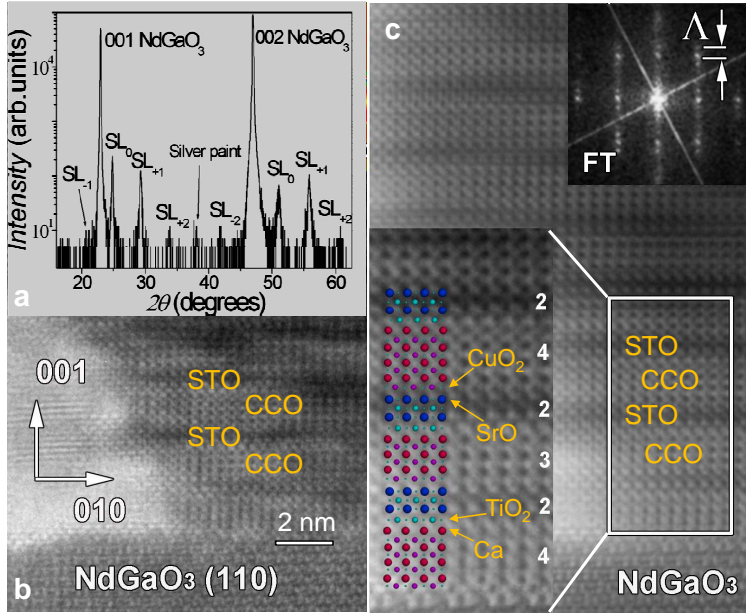}
\end{center}
\caption{(color online) a) XRD pattern (intensity in log scale)
from (CaCuO$_{2}$)\textit{$_{3.5}$}/(SrTiO$_{3}$)\textit{$_{2}$}
SL. SL$_{\pm{i}}$ mark the satellite peaks, around the average
structure peak SL$_{0}$. b) HRTEM image of an identical SL taken
along the [110] orientation of the NdGaO$_{3}$ substrate.  c)
filtered HRTEM image. Bottom inset: enlargement of the area marked
by white rectangle close to substrate with the  atomic model of
STO-CCO sequences (dark blue: Sr; light blue: Ti; small light
blue: O; red: Ca; purple: Cu). Top inset: Fourier Transform
pattern where superlattice spots are clearly visible along [001]
growth direction.} \label{structure}
\end{figure*}

 In Fig.\ref{structure}a the x-ray diffraction (XRD) spectrum
of a superconducting SL (CCO)$_{n}$/(STO)$_{m}$, with nominal
composition $n =$ 3 and $m =$ 2,
 reveals the presence of sharp satellite
peaks around the average structure one, indicating the formation
of a high quality superlattice  with period $\Lambda{}$
$\approx{}$ 19.5(5) \AA{}, \textit{n} = 3.5(5), and \textit{m} =
2.0(5).
To better evaluate the structural quality of CCO/STO SLs, we
recorded a HRTEM image  on an identical sample along the [110]
direction of the NdGaO$_{3}$ substrate. The image
(Fig.\ref{structure}b) shows a heteroepitaxial superlattice film
growth and a series of stacked layers with different contrast,
regularly alternate thicknesses and sharp interfaces. The high
quality of the SL is confirmed by the Fourier Transform (FT)
pattern (inset of Fig.\ref{structure}b), where, beside the intense
reflections associated with the CCO and STO substructures,
satellite lower intensity reflections are observed in the growth
direction (i.e. [001]* direction). These reflections are
associated to the periodic structure generated by the regular
stacking of CCO and STO. The stacking can be described as an
average of 3.5 u.c. of CCO and 2 u.c. of STO, with a superlattice
period $\Lambda{}$ $\approx{}$ 19 \AA{}, in agreement with the XRD
analysis. This means that the number of CCO u.c. varies from 3 to
4. This assignment is confirmed by the analysis of the filtered
HRTEM image (see
 Fig.\ref{structure}c), where it is possible to
precisely determine the layers stacking, as illustrated in the
colored circles model structure in the inset.
\begin{figure}[tbp]
\vspace{0.5cm}
\par
\begin{center}
\includegraphics[width=8 cm]{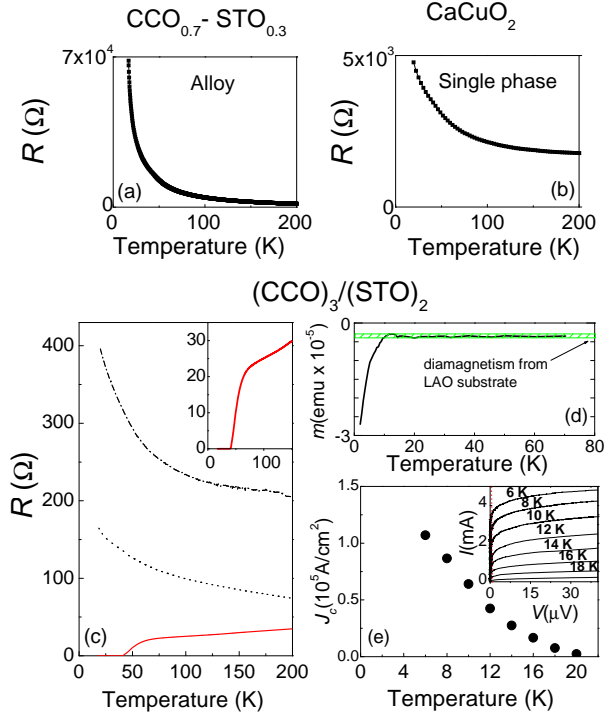}
\end{center}
\caption{(color online) $R(T)$ of the  CCO-STO alloy (panel a) and
the single phase CCO (panel b) grown in the same conditions as the
superconducting SLs. c) $R(T)$ for a SL grown in low oxidizing
atmosphere (dash-dotted line), for one grown in the same condition
and quenched in high (about 1 bar) oxygen pressure (dotted line),
and for a SL grown in a oxygen/ozone mixed atmosphere at higher
pressure and  quenched in high oxygen pressure (full line). Inset:
superconducting transition shown on a reduced temperature range.
d)Magnetic moment as a function of temperature for a (CCO)$_{3}$
/(STO)$_{2}$ SL grown on LaAlO$_{3}$ substrate. The horizontal
green line indicates the value of the temperature independent
diamagnetism of  LAO. e) Critical current density as a function of
temperature for a SL with $T_{c}=$ 22 K after patterning process.
Inset: $I-V$ characteristics at different temperature for the
sample patterned as shown in Fig. \ref{patterned}.
 The vertical dashed line  at V=1$\mu{}$V
corresponds to the criterion adopted for \textit{I$_{c}$}
measurement.} \label{RvsOxy}
\end{figure}

We have observed that the conductivity of (CCO)$_{n}$/(STO)$_{m}$
SLs slightly varies with varying \textit{n} and \textit{m}, but
 increases substantially with increasing the oxidizing power of the
growth atmosphere, till the occurrence of superconductivity.
 In particular,   for a weakly oxidizing growth atmosphere (oxygen
pressure lower than 0.1 mbar) the SLs show always a
semiconductor-like temperature dependence of the resistance
(Fig.~\ref{RvsOxy}c), even if the film is quenched to room
temperature in high (about 1 bar) oxygen pressure. An insulator to
metal transition occurs only when a highly oxidizing growth
atmosphere is used (oxygen plus 12\% ozone at a pressure of about
1 mbar) and the film is rapidly quenched to room temperature at
high oxygen pressure (about 1 bar). Under these conditions, the SL
is metallic, and, in the best case, the resistance goes to zero at
about 40 K (Fig.~\ref{RvsOxy}c). Thus,  strong oxidation is a key
ingredient to obtain superconducting samples.

The occurrence of superconductivity has been also
 characterized by magnetization and critical current density (\textit{J$_{c}$})
 measurements. The temperature dependence of \textit{J$_{c}$} was obtained
 by the measurement, on  suitably patterned samples (see Fig. \ref{patterned}),
  of the \textit{I-V} characteristics
 shown in  Fig. \ref{RvsOxy}e at different temperatures.  To determine \textit{J$_{c}$} at each temperature, a 1  $\mu$V criterion has been
 adopted (vertical dashed line in the inset of Fig. \ref{RvsOxy}e).
   At \textit{T} = 6 K,
 \textit{J$_{c}$} = 1.1x10$^{5}$ A/cm$^{2}$, a value comparable with  other superconducting SLs.  \cite{Gozar}
  The magnetic moment was measured by means of a
commercial SQUID by Quantum Design. Since the NGO substrate is
strongly paramagnetic, \cite{Podlesnyak} and,  at low temperature,
it hinders the detection of the diamagnetic signal from the SL,
then, as a substrate, we used LaAlO$_{3}$ (LAO), which has a small
and temperature independent diamagnetism \cite{Khalid} (see
Fig.\ref{RvsOxy}d). A clear signature of a diamagnetic transition
in (CCO)$_{3}$/(STO)$_{2}$ SL does appear below 12 K. The
\textit{T$_{c}$} in this SL is much depressed compared to similar
SLs grown on NGO, probably because of the larger and opposite
in-plane lattice mismatch between LAO (\textit{a} = 3.78 \AA{})
and CCO (\textit{a} = 3.84 \AA{}).

Once proven the actual occurrence  of a superconducting phase in
CCO/STO SLs by transport and magnetization measurements, it is
important now to point up that the proper doping and the
consequent superconductivity do occur only in the presence of a
layered structure with sharp interfaces. Indeed, we have grown, in
the same strongly oxidizing conditions used for the
superconducting SLs, a CCO-STO alloy (interface free), with 70\%
CCO and 30\% STO, and a pure CCO film: both systems show a
semiconductor-like temperature dependence of the resistance with
no trace of superconductivity (see Fig.~\ref{RvsOxy}a and
\ref{RvsOxy}b). Therefore, the \textit{layered structure} is
needed in order to obtain superconductivity, and we can thus argue
that the excess oxygen atoms enter at the CCO/STO interfaces.
\begin{figure}[tbp]
\vspace{0.5cm}
\par
\begin{center}
\includegraphics[width=6cm]{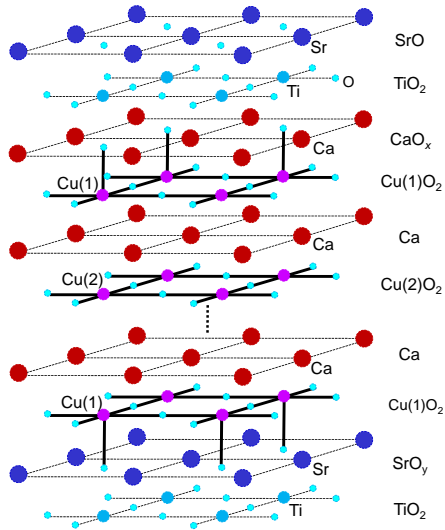}
\end{center}
\caption{(color online) Schematic representation of the layered
structure of the superlattice where the two Cu sites, Cu(1) and
Cu(2) with different oxygen coordination, are show. The thick full
lines represent the coordination of Cu with O.} \label{interface}
\end{figure}

Indeed, there are two kinds of interfaces in CCO/STO SLs (see
Fig.\ref{interface}): CuO$_{2}$-Ca-TiO$_{2}$-SrO and
Ca-CuO$_{2}$-SrO-TiO$_{2}$. Both the interfaces are hybrid between
the perovskite and the IL structure. For instance, the first one
could adopt the IL structure of CaCuO$_{2}$
(CuO$_{2}$-Ca-TiO$_{2}$-SrO) or the perovskite structure of
CaTiO$_{3}$ (CuO$_{2}$-Ca\textit{O}-TiO$_{2}$-SrO), depending on
the oxygen content at the interface Ca plane. Similarly, at the
other interface we can have the IL SrCuO$_{2}$ or the perovskite
SrTiO$_{3}$. Therefore, it is likely that the two interfaces have
CuO$_{2}$-Ca\textit{O}\textit{$_{x}$}-TiO$_{2}$-SrO and
Ca-CuO$_{2}$-Sr\textit{O}\textit{$_{y}$}-TiO$_{2}$ compositions.
Under strongly oxidizing conditions, the overall oxygen content at
the interfaces (\textit{x + y}) could become larger than one,
giving rise to doping.

\begin{figure}[tbp]
\vspace{0.5cm}
\par
\begin{center}
\includegraphics[width=8cm]{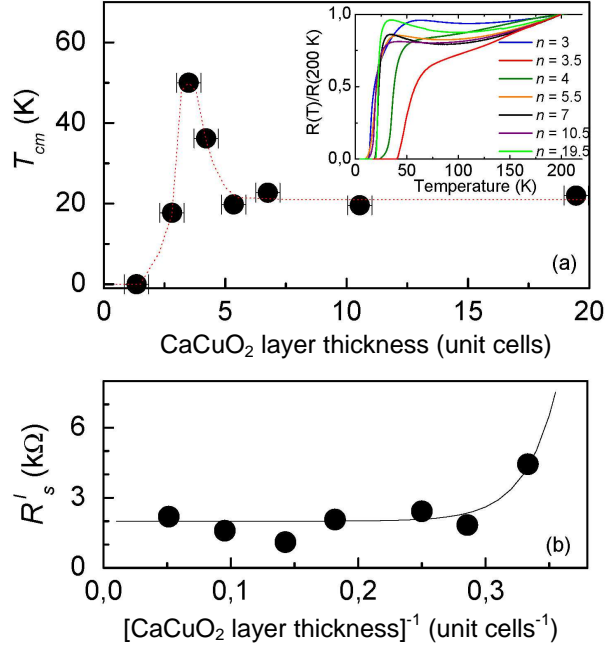}
\end{center}
\caption{(color online) a) \textit{T$_{cm}$}, defined as the
midpoint of the resistive transitions shown in the inset, as a
function of the number $n$ of unit cells  of the CCO. The dotted
line is a guide for the eye. Inset: \textit{R(T)} normalized at $T
=$ 200 K for the same SLs as in the main panel. b)sheet resistance
at 80 K of one CCO layer \textit{R$^{l}$$_{s}$} as a function of
the inverse of the number of CCO unit cells. Full line is a guide
to the eye.} \label{Rs}
\end{figure}

Since we expected that the doping  occurs at the interface, we
investigated how far from the interfaces superconductivity
extends. To this aim we prepared a series of SLs, where the number
of STO u.c. is kept fixed at \textit{m} = 2 and the number of CCO
u.c. is varied from \textit{n} $\approx{}$ 1.5 to \textit{n}
$\approx{}$ 20 (see section II.A). Superconductivity shows up when
\textit{n} becomes larger than 2, and \textit{T$_{cm}$}, defined
as the transition midpoint in the \textit{R(T)} curves (inset to
Fig.\ref{Rs}a), reaches the maximum value for \textit{n} between 3
and 4 (Fig.~\ref{Rs}a). For higher values of \textit{n},\textit{
T$_{cm}$} decreases and, for \textit{n} $>$ 5 remains almost
constant up to the thickest sample (\textit{n} $\approx{}$ 20).
This behavior suggests that superconductivity is confined within
few unit cells from the interface. To further confirm this
conclusion, we analyzed the sheet resistance \textit{R$_{s}$}
multiplied by the number \textit{M} of CCO layers within the
superlattice (i.e., the number of repetitions of the
CCO$_n$/STO$_m$ supercell): \textit{R$^{l}$$_{s}$} = \textit{M} x
\textit{R$_{s}$}, which is thus the sheet resistance per CCO
layer. If the carriers were uniformly distributed over the whole
CCO layer, \textit{R$^{l}$$_{s}$} would depend linearly on the
inverse of the number \textit{n }of CCO unit cells in the layer,
extrapolating to \textit{R$^{l}$$_{s}$} = 0 at \textit{n$^{-1}$} =
0. But this is not the case, since, as shown in Fig.~\ref{Rs}b,
\textit{R$^{l}$$_{s}$} does not appreciably vary with the CCO
thickness, except for the SL with the thinnest CCO layer.

Given that superconductivity in CCO/STO SLs deals with the
interface properties, it is  worth mentioning here that at the
CCO/STO interface a large polar discontinuity is present, since
the planes TiO$_2$ and SrO of STO are neutral, whereas the planes
Ca and CuO$_2$ of the CCO are charged (+2\textit{e} and
-2\textit{e} per unit cell, respectively). Therefore, some sort of
interface reconstruction should occur in order to suppress this
huge built in electrostatic potential. Most likely, a pure ionic
mechanism involving displacement of oxygen ions, similar to the
case of the interface between SrO terminated STO and LAO,
\cite{Nakagawa} is at work in CCO/STO SLs, due to the hybrid
(perovskite/infinite layer) interfaces, where the presence of a
variable oxygen content is allowed.

We have up to now shown that: i) it is possible to synthesize
novel cuprate/titanate heterostructures with high structural
quality; ii) a high $T_c$ superconducting phase does actually
occur in these systems; iii) sharp interfaces and strongly
oxidizing growth conditions are needed to obtain
superconductivity; iv) (super)conductivity is confined within few
unit cells from the interfaces. Therefore, the engineering of the
 synthetic heterostructure CCO/STO led to a new phase of
electron matter, i.e., high $T_c$ superconductivity, which does
not exist in the single insulating constituent oxides, being
exclusively a consequence of a process occurring at the interface
between  CCO and STO.

To reach a deeper knowledge of the process operating at the
interface, which involves the presence of excess oxygen ions, we
performed additional key experiments, whose results are
illustrated in the following.

\begin{figure}[tbp]
\vspace{0.5cm}
\par
\begin{center}
\includegraphics[width=7cm]{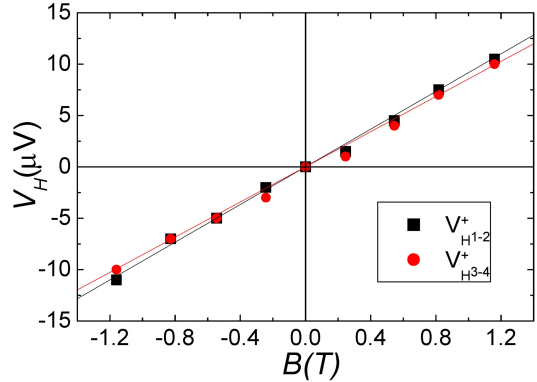}
\end{center}
\caption{(color online) Hall voltage \textit{V$_{H}$} vs. external
magnetic field \textit{B} for a given direction of the bias
current fixed at \textit{I} $=$ 0.1 mA. Square and circles refers
to \textit{V$_{H}$} values acquired between terminals 1-2 and
terminals 3-4, respectively, as indicated in the picture of the
patterned sample (Fig. \ref{patterned}. The straight lines are
linear fit to the data with the slope proportional to
\textit{R$_{H}$}.} \label{Hall}
\end{figure}
Hall voltage (\textit{V$_{H}$}) measurements were performed on
suitably patterned samples (see Fig. \ref{patterned}).
\textit{V}$_{H}$ was measured at room temperature between the
opposite voltage contacts biasing the sample with a 100 $\mu{}$A
dc current and applying an external magnetic field \textit{B} up
to 1.2 T. In Fig. \ref{Hall}, \textit{V$_{H}$} as a function of the
external magnetic field \textit{B} is
 shown for both the voltage configurations and for one direction of the current. For each configuration a
 Hall constant \textit{R$_H$} is obtained by the slope of the best fit straight line through the experimental data.
 By the average of the calculated slopes of the different experimental data set
 (two opposite voltage terminals and two current directions), the value \textit{R$_{H}$ }= 3.4x10$^{-3}$cm$^{3}$/C
 was obtained.
The positive sign of \textit{R$_{H}$} gives a strong indication of
the hole nature of the majority charge carriers in these
materials. The calculated charge density is \textit{n$_{H}$ }= 1.8
x 10$^{21}$ cm$^{-3}$, which results comparable with that observed
in other cuprate superconductors and cuprate based SLs.
\cite{Affronte}

One of the original characteristics of the CCO/STO SLs, with
respect to other previously reported high $T_c$ HSs,
\cite{Bales1,Aruta,Gozar} is that in CCO/STO Cu is present only in
one of the two constituent blocks. In this case a block selective
study is possible by resonant spectroscopic techniques. We thus
additionally performed x-ray absorption spectroscopy
measurements \cite{Aruta,Fink}.
Indeed, XAS is a well established synchrotron based technique
providing chemical and site selective information on the
electronic states close to the Fermi level. In highly
 correlated 3d transition metal systems $L_{2,3}$ edge XAS (mainly 2p$\rightarrow{}$3d transitions) can reveal the
 symmetry of unoccupied 3d states and distinguish among sites with different valence and coordination.
 XAS spectra were collected at 5 K at the Cu and Ti
\textit{L}$_{2,3}$-edges \cite{Groot}on a superconducting
(CCO)$_{3}$/(STO)$_{2}$ SL with \textit{T$_{c}$} = 25 K, and on an
identical SL, but \textit{non-superconducting}, since grown with
no ozone and with no quenching at high oxygen pressure.  A linear
background was
 fitted to the pre-edge region of the $L_3$ edge and subtracted from the spectra, which are then normalized
 to the edge jump set to unity above the $L_2$ edge.
\begin{figure}[tbp]
\vspace{0.5cm}
\par
\begin{center}
\includegraphics[width=9cm]{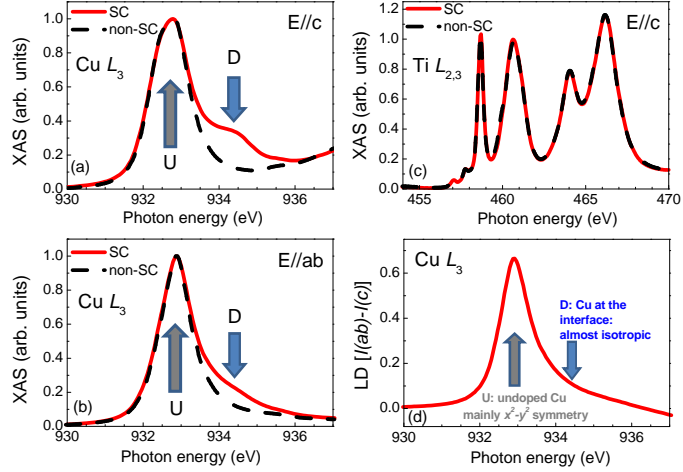}
\end{center}
\caption{(color online) a,b: Cu L$_{3}$-edge for a superconducting
(SC) (full line) and non-superconducting (dashed line)
(CCO)$_{3}$/(STO)$_{2}$ SL. All measurements are performed at $T
=$ 5 K and the spectra are normalized to the maximum height
intensity of the L$_{3}$ edge.  c) the two multiplets Ti
\textit{L}$_{2}$- and Ti \textit{L}$_{3}$ - edges for SC and
non-SC sample. d)linear dichroism for the SC  sample.} \label{XAS}
\end{figure}
 Figures \ref{XAS}a and \ref{XAS}b show the normalized XAS
spectra taken at the Cu \textit{L}$_{3}$-edge (Cu
2\textit{p}$\rightarrow{}$Cu 3\textit{d} electron transition
energy) for the two samples, with the electric field of the
incident radiation parallel to the c-axis (\textbf{E}//c) and to
the ab-plane (\textbf{E}//ab), respectively.
The \textit{L}$_{3}$ peak at 932.8 eV, indicated with U in
Fig.\ref{XAS}, is associated to the process in which the
3\textit{d}$^{9}$ ground state becomes \uline{c}3\textit{d}$^{10}$
(3\textit{d}$^{9}$$\rightarrow{}$\uline{c}3\textit{d}$^{10}$),
where \uline{c} indicates a Cu 2p core hole. This peak is assigned
to the absorption by an undoped Cu site.
\cite{Aruta,Bianconi,Sarma}
 The feature indicated with D
at about 1.5 eV
above the main peak U is associated to the process
3\textit{d}$^{9}$\uline{L}$\rightarrow{}$\uline{c}3\textit{d}$^{10}$\uline{L},
where \uline{L} indicates  an additional oxygen ligand hole
arising from Cu 3\textit{d} -- O 2\textit{p} hybridization, mainly
with an O 2\textit{p} character. \cite{Aruta,Bianconi,Fink} Thus,
this feature is assigned to the absorption by doping
holes.\cite{Aruta,Bianconi,Sarma}
 The major differences between the spectra of
superconducting and non-superconducting SL are observed in the
region around D. In both polarizations, this region acquires
spectral weight in the case of superconducting SL, as a clear
indication of increased holes concentration delocalized on the
in-plane and out-of-plane O 2\textit{p} bands.
 We note here that the presence of
out-of-plane ligand holes has been already observed in some HTS
\cite{Singhal,Srivastava} and the debate about their role in
superconductivity is still open.
On the other hand, at the Ti \textit{L}$_{2,3}$ -- edges, no
relevant changes are observed between the superconducting and the
non-superconducting samples (Fig.\ref{XAS}c). According to ref.
\onlinecite{Verbeeck}, in case of Ti$^{3+}$ doping, an increase of
the intensity would be expected in the region between the peak at
464 eV, associated to the 3\textit{d}-t$_{2g}$ final states, and
the peak at 466 eV, associated to the 3\textit{d}-e$_{g}$ final
state. This increase, in our case, does not occur. Therefore, not
only the valence, but also the crystal-field in the Ti environment
\cite{Verbeeck} are not substantially affected by the strongly
oxidizing conditions used to obtain the superconducting sample.

Thus, by a direct observation of the XAS spectra we have confirmed
that the charge carriers are holes and established that they are
located in the CCO block.

With the help of the Cu-\textit{L}$_{3}$-edge linear dichroism
(LD) showed in Fig. \ref{XAS}d, a more detailed comprehension of the
XAS spectral features can be obtained.
The XAS LD (Fig.\ref{XAS}d) is the difference between the XAS
contributions from the two polarizations: LD =
\textit{I}(\textbf{E}//ab)- \textit{I}(\textbf{E}//c), where both
spectra have been normalized to the value of the \textit{L}$_{3}$
intensity of the \textbf{E}//ab spectrum. LD is thus a measure of
the anisotropy of the environment around selected ions.
In the SLs CCO/STO we
can distinguish two distinct Cu sites (see Fig.\ref{interface}):
 i) Cu(1) sites at the CuO$_2$ planes closest to  STO (interface
 sites), which can have apical oxygen;
ii) Cu(2) sites at all the other CuO$_2$ planes, all of them with
purely planar oxygen coordination. Upon doping (strong
oxidization), the majority of Cu(1) sites are expected to have an
apical oxygen  because of additional oxygen atoms entering the
interface planes.  Since the c-axis of CCO is about 3.19 \AA, the
out-of-plane Cu(1)-O distance (see Fig. \ref{interface}) is not
much different from the in -plane one, as it has been also shown
in the doped CCO with planar defects. \cite{Zhang} Therefore, an
isotropic
 orbital population is expected at Cu(1) sites.
Recalling that the main peak U is ascribed to the absorption by
undoped Cu sites and the region D to the absorption by delocalized
doping holes, we can assign D, which is just weakly anisotropic
(low LD), to the  doped  Cu(1) sites hybridized with both in-plane
(holes with 3d$_{x2-y2}$ character) and out-of-plane (holes with
3d$_{3z2-r2}$ character) oxygen orbitals.
On the other hand, the presence in the XAS spectra  of spectral
weight with large LD in the region about 1 eV above U(see Fig.
\ref{interface}d),  can be ascribed to the absorption by doping
holes with mostly in-plane 3d$_{x2-y2}$ character \cite{Chen} at
the Cu(2) sites, which have purely planar oxygen coordination. We
can thus conclude that the doping holes, although mainly confined
at the Cu(1) sites, can get transferred to the neighboring
Cu(2)O$_2$ planes, giving rise to an interface-like
superconductivity.

\section{Summary}

 In summary, we synthesized the
superconducting heterostructure
(CaCuO$_{2}$)$_n$/(SrTiO$_{3}$)$_m$, with maximum $T_c =$ 40 K, by
alternating an insulating IL cuprate (CCO) and a copper-free wide
gap semiconductor (STO). The charge reservoir role is played by
the  interfaces, due to the peculiar structural and electronic
properties of the constituent oxides, which allow extra oxygen
ions entering the interface planes. The doping holes, from the
interface layers, are injected in the inner CuO$_2$ planes close
to the interface, whereas STO does not contribute to transport.
This work could motivate the search for other cuprate/non-cuprate
synthetic heterostructures, in which the CuO$_2$ plane properties
(in-plane Cu-O distance, buckling), the charge reservoir
characteristics, and the out-of plane unit cell size can be
independently controlled. This possibility  may lead to improved
superconducting properties of the HSs and to unveil open questions
concerning superconductivity in cuprates.

W.P. and O.I.L. are grateful to Dr. A. Pautrat for helpful
discussion and to M. L. Gouleuf for the preparation of the samples
for HRTEM measurements. D.D.C. thanks A. Maisuradze for support
during magnetization measurement. C.A. and D.D.C. thanks J.
Zeghenagen for fruitful discussions. This work was partly
supported by the Italian MIUR (Grant No. PRIN-20094W2LAY, "Ordine
orbitale e di spin nelle eterostrutture di cuprati e manganiti").

\end{document}